\documentstyle[aps,preprint,prb,epsf]{revtex} 

\include{epsf}  
\begin{document} \baselineskip 24pt

\title{\Large \bf Scaling Analysis and Evolution Equation \\ of the North
Atlantic Oscillation Index Fluctuations}

\author{C. Collette$(*)$ and M. Ausloos \\SUPRATECS, B5, Sart Tilman, B-4000
Li$\grave e$ge, Belgium \\ (*) present address: Active Structures Laboratory,
Universit\'e Libre de Bruxelles, \\ ULB - CP165/42 av. F.D. Roosevelt, 50 B-1050
Brussels, Belgium}

\date{\today}

\maketitle

\begin{abstract} The North Atlantic Oscillation (NAO) monthly index is studied
from 1825 till 2002 in order to identify the scaling ranges of its fluctuations
upon different delay times and to find out whether or not it can be regarded as a
Markov process. A Hurst rescaled range analysis and a detrended fluctuation
analysis both indicate the existence of weakly persistent long range time
correlations for the whole scaling range and time span hereby studied. Such
correlations are similar to Brownian fluctuations. The Fokker-Planck equation is
derived and Kramers-Moyal coefficients estimated from the data. They are
interpreted in terms of a drift and a diffusion coefficient as in fluid
mechanics. All partial distribution functions of the NAO monthly index
fluctuations have a form close to a Gaussian, for all time lags, in agreement
with the findings of the scaling analyses. This indicates the lack of predictive
power of the present NAO monthly index. Yet there are some deviations for large
(and thus rare) events. Whence suggestions for other measurements are made if
some improved predictability of the weather/climate in the North Atlantic is of
interest. The subsequent Langevin equation of the NAO signal fluctuations is
explicitly written in terms of the diffusion and drift parameters, and a
characteristic time scale for these is given in appendix. \end{abstract}

\vskip 0.15cm


\newpage

\section{Introduction}

In order to establish a sound understanding for any scientific phenomenon, one
has to record numerical data and from the latter to obtain laws which can be next
derived theoretically from so called first principle models. The so called
inverse model method, starting from raw data and using statistical analysis as a
first step, is of great interest since it is model free. Some difficulty arises
in particular in nonlinear dynamical systems because of the need to sort out
noise from both chaos and deterministic
components\cite{BhattacharyaandKanjilal2000,Provenzaleetal1992}. Whence to
extract meaningful model-free dynamical equations from chaotic-like data is an
enormous challenge\cite{RowlandsandSprott1992}. Practically one is often led to
empirical relationships. This is often the case in the meteorology/climatology
field where there is a widely mixed set of various (sometimes) unknown
influences, over different time and space scales. Often the fast variations are
taken as noise terms in some sort of Langevin
equation(s)\cite{PenlandandMatrosova1998,PenlandandSardeshmukh1995} as for the
{\it el Ni$\tilde{n}$o} Southern Oscillation Index (SOI).

In order to quantify weather and climate events in Europe and report large-scale
variability an index has been imagined\cite{Marshall et al. 2001} the so called
North Atlantic Oscillation (NAO) index ($http://www.ldeo.columbia.edu/NAO/$;
$http://www.met.rdg.ac.uk/cag/NAO/$; $http://www.cru.uea.ac.uk/cru/info/nao/$;

$http://www-bprc.mps.ohio-state.edu/gpl/NAO/Naobibliography.htm$). It is the
normalized sea level pressure (SLP) difference between a station at Ponta
Delgada, Azores and one at Akureyri, Iceland.

Since about the mid-50's the NAO index was trending from negative to positive
values, but is mostly positive since 1980, a variation attributed to global
warming. It is thought\cite{Hurrell et al. 2001} that the influence of slow
changes in the ocean and in the greenhouse gases maybe picked up as the
fundamental causes of a prolonged (upward) trend.

Until a few years ago, the NAO was not receiving intense attention\cite{Hurrell
et al. 2001}, because it was thought that its phase and amplitude were rather
unpredictable, because both involve many (time and space) scales which are often
intrinsic to chaotic behavior; see also reviews\cite{Palmer (2000),Greatbatch
2000,Wanner et al. (2001)}. Yet, evidence has been presented that NAO exhibits
'long-range' dependence having winter values residually correlated over many
years, with short-term 2-5 year variations and decadal trends. Note that
Wallace\cite{Wallace (2000)} has argued that the NAO is a local expression for a
Northern Annular Mode (NAM), also called Arctic Oscillation\cite{Ambaum et al.
(2001),Stephenson et al. (2000)}.

In view of the above it seems of pertinent interest to consider again the NAO and
adopt specific data analysis techniques when searching for scaling ranges and
stochasticity features. We start with the Hurst (R/S) method \cite{Hurst
1951,Turcotte 1997} followed by a detrended fluctuation analysis (DFA)\cite{Azbel
1995,Hu et al. 2001} of the monthly averaged NAO signal. Such tests supplement
classical analyses based on frequency spectra\cite{Pelletier 1997,Talkner and
Weber 2000,Fernandez et al. 2003} which are debatable due to the non stationarity
of the data. Interestingly the data histogram have so called fat tails,
resembling the L\'evy flights, signatures of $self-organizing$ systems, today
emerging in many areas of physics as those mentioned here above. Again these
facts seem to exclude low dimensional chaos but support the conjecture of Markov
dynamics for atmospheric evolution, as already suggested in fact many years
ago\cite{Palmer (2000),Hasselmann (1976),Frankignoul and Hasselmann(1977),Leith
(1978)}.

In the following, we adopt a Markov assumption in order to derive the FPE and to
write down the Chapman-Kolmogorov equation for the conditional probability of the
increments $\Delta x$ (of the NAO index) over different time intervals $\Delta
t$. This leads to a numerical derivation of the Kramers-Moyal coefficients which
are the moments of such probability distributions. Up to the second moments, this
leads to the $diffusion$ and $drift$ coefficients appearing in the Fokker-Planck
equation (FPE) and are basic to the Langevin equation\cite{Risken 1984}. It will
be noticed that the analytical form of both drift $D^{(1)}$ and diffusion
$D^{(2)}$ coefficients are simple. It will be found that the $experimental$
probability density functions (pdf) have all a Gaussian form when excluding the
(rare) large (so called extreme) events.

The methods, applied in this paper, are briefly explained in Sect. 2.2 and 2.3 :
(1) the rescaled range analysis and (2) the detrended fluctuation analysis. In
all cases results are tested against surrogate or shortened data for error bar
evaluation. Both methods lead to an exponent characteristic of the classical
random walker position fluctuation correlations. Next, in Sect. 3, we   examine
how to describe the statistical evolution of increments for different time scales
i.e. establishing a Fokker-Planck equation within a Markov process assumption. A
few comments pertain to considerations on NA weather causes and predictions in a
discussion and conclusion sections.

\section{Data and theoretical analysis}

\subsection{Data}

The monthly averaged NAO index (available on the web sites
$http://www.cru.uea.ac.uk/$ $ftpdata/nao.dat$ or
$http://www.cru.uea.ac.uk/cru/data/nao.htm$ and updated at
$http://www.cru.uea.ac.uk/$$\tilde{}$$timo/projpages/nao_{-}update.htm$), i.e. as
the normalized sea level pressure (SLP) difference between a station at Ponta
Delgada, Azores (26$^{\circ}$W,38$^{\circ}$N) and one at Akureyri, Iceland
(18$^{\circ}$W,66$^{\circ}$N) is represented on Fig.1 from January 1825 to
November 2002 (2135 points).\footnote{ For completeness let us point out that
early instrumental or paleoclimatic data can be used to extend the North Atlantic
Oscillation index back to 1823 or even 1675\cite{Jones et al. 1997,Jones et al.
1998,Luterbacher et al. 1999,Schmutz et al. (2000)}.}

It is a standard procedure that in order to reduce spurious noise effects, the
study is performed on the $integrated$ series (Fig.2). Such values can be
interpreted as mimicking the successive positions of a random walker
\cite{Turcotte 1997}. The amplitude correlations should allow us below to
understand the drift and diffusion process (as that of a walker).

\subsection{The rescaled range analysis}

Introduced by Hurst\cite{Hurst 1951,Malamud and Turcotte 1999}, the rescaled
range analysis method computes a ratio $R/S$ defined as follows. The time series
$X=\left\{ x_{t},t=1,...,N\right\} $ is divided into $l$\ intervals of equal
length $n$. In the $k^{th}$ box, ($k=1,...,l$), there are $n$ elements,
$X_{j}^{(k)}(n)=\left\{ x_{j},j=(k-1)n+1,...,(k-1)n+n (\equiv kn)\right\} $. The
local fluctuation at point $j$ in the $k^{th}$-box, i.e. $\left(
x_{j}^{(k)}-\left\langle x\right\rangle _{n}^{(k)}\right) $ is calculated as the
deviation from the mean $\left\langle x\right\rangle _{n}^{(k)}=\frac{1}{n}
\sum_{j=1}^{n}x_{j}^{(k)}$, in that $k^{th}$-box. The $cumulative$ departure
$Y_{m}^{(k)}(n)$ up to the $m^{th}$ point in the $k^{th}$-box (of size $n$) is
next calculated

\begin{equation} Y_{m}^{(k)}(n)=\sum_{j=1}^{m}\left( x_{j}^{(k)}-\left\langle
x\right\rangle _{n}^{(k)}\right) =(\sum_{j=1}^{m}x_{j}^{(k)})-m\left\langle
x\right\rangle _{n}^{(k)} \label{ers1} \end{equation} for $m=1,..., n$ and in all
$k$ boxes and where $\left\langle x\right\rangle^{(k)} _{n}=\frac{1}{n}
\sum_{j=1}^{n}x_{j}^{(k)}.$ The rescaled range function is defined by

\begin{equation} \frac{R^{(k)}}{S^{(k)}}(n)=\frac{\max_{1\leq m\leq n}\left(
Y_{m}^{(k)}(n)\right) -\min_{1\leq m\leq n}\left( Y_{m}^{(k)}(n)\right) }{\sqrt{
\left( \frac{1}{n}\right) \sum_{j=1}^{n}\left( x_{j}^{(k)}-\left\langle
x\right\rangle _{n}^{(k)}\right)^{2}}}\quad \quad \quad k=1,...,l . \label{ers}
\end{equation}

The average of the rescaled range in all boxes with an equal size $n$ is next
obtained and denoted by $<R/S>$. The above computation is then repeated for
different values of $n$ to provide a relationship between $<R/S>$ and $n$, -
which is expected to be a power law $<R/S>\simeq n^{H }$ if some scaling range
and law exist; $H$ is called the Hurst exponent. If $H =0.5$, the signal is
uncorrelated (white noise); the ''walk'' is like a Brownian motion. If $H <0.5$,
the signal is anticorrelated (blue noise); if $H >0.5$, there are positive
correlations in the signal (red noise). From Fig. 3 it is found that $H =0.55 \pm
0.02$ for the NAO index variations from 10 to 300~months (about 25~years). The
departure from strict linearity on a log-log plot is usually attributed to too
small box sizes, or to some periodic mode not finally taken into account through
the assumed constant base line in Eq. (1); see discussions in references quoted
here above and in the introduction. In order to test the robustness of the result
we have checked the scaling properties of NAO index data series that are shorter
than the original one by 5\% (107 data points). In both cases when we delete the
first 107 data points or the last 107 data points, the R/S analysis exponent has
roughly the same value. The same goes true (not shown) for surrogate data series,
i.e. when amplitudes are randomly displaced or multiplied by a random sign. It is
certain that the $H$ value near 0.55 indicates a weak deviation of the signal
from Brownian motion.

\subsection{The detrended fluctuation analysis}

The detrended fluctuation analysis method has been recently much used in the
meteorological field\cite{Talkner and Weber 2000,Koscielny-Bunde et al.
1998,Ivanova and Ausloos 1999,Ivanova et al. 2000}. The method has the advantages
of avoiding (seasonal-like) trends and non stationarity effects, intrinsic to the
finite size of the data. The method consists in dividing the time series
$X=\left\{ x_{i},i=1,...,N\right\} $ into $l$\ boxes of equal size $n$. In the
k$^{th}$ box, the $cumulative$ sum $Y_{m}^{(k)}$ can be calculated as above, in
the so called the first order DFA. Let $ Y_{fit,i}^{(k)}(n),$ be the best linear
fit to the data in the k$^{th}$ box. The detrended fluctuation function is next
calculated, i.e. dropping the $(n)$, $\phi
_{i}^{(k)}=Y_{i}^{(k)}-Y_{fit,i}^{(k)}.$ The root mean square fluctuation is then
given by

\begin{equation} F(n)=\sqrt{\frac{1}{N}\sum_{i=1}^{N}\left[ {\phi }_{i}\right]
^{2}}. \label{edfa} \end{equation} If the values of the time series are
correlated, there is a power-law relationship between $F(n)$ and $n$: $F(n)\simeq
n^{\alpha }$. A departure from linearity on a log-log plot, and the existence of
crossovers (hereby one occurs near 240 months, see Fig.4) has been
discussed\cite{Hu et al. 2001}. Fig. 4 shows that $\alpha =0.54\pm 0.02$ below
240 months, for NAO index variations. Again the robustness of the result is
confirmed by analysing shorter or surrogate data series. Various
considerations\cite{Turcotte 1997} indicate that $\alpha $ should be equal to
$H$. Thus we can conclude that the value of the scaling exponent $\alpha$ is
roughly the same as the one obtained within the rescaled range analysis. Notice
that the DFA method, is clearly leading to extend the scaling properties of the
NAO index toward smaller scales, less than 10 months as is found for the R/S
analysis.

The above findings confirm the existence of non trivial correlations (since
$\alpha \neq 0.5$, even though it is close to 0.50) within precise interval time
ranges. They point out to the existence of physical phenomena described as
fractional Brownian motions (Mandelbrot 1982) thus with a fractal-like hierarchy
of $time$ scales. The result of $\alpha$ values larger than 0.5 can be
interpreted again through a persistence effect in the fluctuations\cite{Turcotte
1997,Malamud and Turcotte 1999}.

\subsection{The Fokker-Planck equation}

In view of the above it is of interest to search whether these weak and
persistent correlations can be found through the solution of a phenomenological
{\it evolution equation}, like the Fokker-Planck equation\cite{Pecseli2000}. Thus
we focus on the variations $\Delta x$ of the elements of the NAO series and the
more so on their distribution in time. In order to do so we follow the method of
Friedrich et al.\cite{Friedrich et al. 2000} and reproduce it \textit{almost in extenso} here below, for the technique is not necessarily familiar to most
readers.

ln order to characterize the statistics of NAO changes, increments $\Delta
x_{1}$, $\Delta x_{2}$, ... for delay times $\Delta t_{1}$, $\Delta t_{2}$, ...
at the same time $t$ are considered. This leads to a set of $p(\Delta
x_{i},\Delta t_{i}$). Next the joint probability density functions are evaluated
for various time delays $\Delta t_{1}$ $>$ $ \Delta t_{2}$ $>$ $\Delta t_{3}$ $>$
... directly from the given data set, e.g. $p(\Delta x_{1},\Delta t_{1};\Delta
x_{2},\Delta t_{2})$. Of course if two increments i.e. $\Delta x_{1}$ and $\Delta
x_{2}$ are statistically independent, the joint pdf should factorize into a
product of two probability density functions:

\begin{equation} p(\Delta x_{1},\Delta t_{1};\Delta x_{2},\Delta t_{2})=p(\Delta
x_{1},\Delta t_{1})p(\Delta x_{2},\Delta t_{2}). \end{equation} leading to an
isotropic single hill landscape in the $\Delta x_{1}, \Delta x_{2}$ plane.

A complete characterization of the statistical properties of the data set in
general requires the evaluation of joint pdf's $p^{N}$($\Delta x_{1}$,$ \Delta
t_{1}$;...;$\Delta x_{N}$,$\Delta t_{N}$) depending on $N$ variables (for
arbitrarily large $N$). In the case of a Markov process (a process without memory
but governed by probabilistic conditions), an important simplification arises:
The $N$-point pdf $p^{N}$ is generated by the mere product of conditional
probabilities $p ( \Delta x_{i+l},\Delta t_{i+l}|\Delta x_{i},\Delta t_{i}) $
itself equal to $ p(\Delta x_{i+l},\Delta t_{i+l};\Delta x_{i},\Delta
t_{i})/p(\Delta x_{i}, \Delta t_{i})$ for $i = 1,...,N-1$. The conditional
probability is given by the probability of finding $\Delta x_{i+1}$ values for
fixed $\Delta x_{i}$. As a necessary condition of Markov processes, the
Chapman-Kolmogorov equation in its integral form reads \begin{equation} p(\Delta
x_{2},\Delta t_{2}|\Delta x_{1},\Delta t_{1})=\int d(\Delta x_{i})p(\Delta
x_{2},\Delta t_{2}|\Delta x_{i},\Delta t_{i})p(\Delta x_{i},\Delta t_{i}|\Delta
x_{1},\Delta t_{1}) \end{equation} and should hold for any value of $\Delta
t_{i}$, with $\Delta t_{2}$ $<$ $\Delta t_{i}$ $<$ $\Delta t_{1}$; see Appendix A
for a discussion in particular concerning large (and thus rare) events. As is
well known, such a Chapman-Kolmogorov equation yields an evolution equation for
the change of the conditional distribution functions $p$ ($\Delta x,\Delta
t|\Delta x_{1},\Delta t_{1}$) and $p$($\Delta x$,$\Delta t$ ) across the scales
$\Delta t$. The Chapman-Kolmogorov equation when formulated in $differential$
form yields a master equation, which can take the form of a Fokker-P1anck
equation\cite{Friedrich et al. 2000,Hanggi and Thomas 1982,Gardiner 1983,Risken
1984}. It is useful to use reduced time units, like $\tau=log_2(16/\Delta
t)$,\footnote{Why ''16'' is chosen to be the normalizing value will be made clear
below, but it has obviously not much effect at this stage.}

\begin{equation} \frac{d}{d\tau}p(\Delta x,\tau )=[-\frac{\partial }{\partial
\Delta x} D^{(1)}(\Delta x,\tau )+\frac{\partial }{\partial ^{2}\Delta x^{2}}
D^{(2)}(\Delta x,\tau )]p(\Delta x,\tau ) \label{efp} \end{equation} in terms of
a drift $D^{(1)}$($\Delta x$,$\tau $) and a diffusion coefficient
$D^{(2)}$($\Delta x$,$\tau $) (thus values of $\tau$ represent $\Delta t_{i}$,
$i=1,...$.) \ Their functional dependence can be estimated directly from the
moments $M^{(k)}$ (known as Kramers-Moyal coefficients) of the conditional
probability distributions:

\begin{equation} M^{(k)}=\frac{1}{\Delta \tau }\int d\Delta x^{^{\prime }}(\Delta
x^{^{\prime }}-\Delta x)^{k}p(\Delta x^{^{\prime }},\tau +\Delta \tau |\Delta
x,\tau ) \end{equation} for different small $\Delta \tau $'s, such that for
$\Delta \tau \rightarrow 0$,

\begin{equation} D^{(k)}(\Delta x,\tau ) \simeq \frac{1}{k!}{lim_{\Delta \tau \rightarrow 0}}   M^{(k)}. \end{equation} After calculating such moments from the
conditional probabilities, we find (Fig. 5) that the coefficient $M^{(1)}$ shows
a linear dependence for small $\Delta x$, while $ M^{(2)}$ can be approximated by
a polynomial of degree two in $\Delta x$. Therefore the type of fluctuation
probability drift term $D^{(1)}$ is well approximated by a linear function of $
\Delta x$, whereas the diffusion term $D^{(2)}$ follows a function quadratic in
$\Delta x$. For very large values of $\Delta x$ the statistics becomes poorer and
the uncertainty increases.

From a careful analysis of the data based on the functional dependences of
$M^{(1)}$ and $M^{(2)}$ (Fig. 5 (a-b)), the following approximations hold true:

\begin{equation} \left\{ \begin{array}{ll} D^{(1)}=-0.52\Delta x + 0.04 \qquad &
for \mid \Delta x \mid <5 \,(NAO~ units)\\ D^{(2)}=\frac{1}{2}\left( 0.21\Delta
x^{2}-0.02\Delta x+4.2\right) \qquad & for \mid \Delta x \mid < 5 \,(NAO~ units)
\end{array} \right. \label{em12nao} \end{equation} Notice the range of validity
of the simple analytical forms, thus the limit found for what would be
called\cite{Grasso and Sornette 1998} ''extreme events'' or ''outliers''. Also
observe that except for the independent term, $2 D^{(2)}\simeq (D^{(1)})^2$, the
strict equal sign being a request for indicating an absolute lack of
intermittency in turbulence\cite{Kolmogorov 1941a,Kolmogorov1941b,Parisi and
Frisch 1985,Frisch 1995}.

The FPE for the distribution function is known to be equivalent to a Langevin
equation for the variable, i.e. $\Delta x$ here, within the Ito
interpretation\cite{Risken 1984,Pecseli2000}

\begin{equation} \frac {d}{d\tau} \Delta x(\tau) = D^{(1)}(\Delta x(\tau),\tau) +
\eta(\tau) \sqrt {{D^{(2)} (\Delta x(\tau),\tau)}}, \end{equation}

\noindent where $\eta(\tau)$ is a fluctuating $\delta$-correlated force with
Gaussian statistics, i.e. $<$ $\eta(\tau)$ $\eta(\tau')$$>$ = 2$ \delta (\tau
-\tau')$. An interpretation of the analogy between these $drift$ and $diffusion$
coefficients and those usually employed in fluid mechanics is given in Appendix.
It may be worthwhile to emphasize here that (i) a negative slope value for
$D^{(1)}$ indicates a sort of restoring or damping force for the evolution of
$\Delta x$; (ii) the observed quadratic dependence of the diffusion term
$D^{(2)}$ is essential for the logarithmic scaling of the intermittence parameter
in turbulence\cite{Parisi and Frisch 1985,Frisch 1995}.

Using those analytical expressions of the empirically derived Kramers-Moyal
coefficients, Eq.(\ref{efp}) can now be integrated, thereby leading to a  test of
the Markov process assumption. For the above change of variables, $\tau = log_2
(16/\Delta t)$, we take the observed distribution of the time series for $\Delta
t=16$~months thus at ''large time'', as the initial condition for the
integration, - considering that there is no propagation of anomalous correlation
to be expected after such a time lag. Indeed the scaling ranges of $H$ and
$\alpha$ found here above (Figs. 3 and 4) indicate that the correlations are
quasi identical for $\Delta t=16$~months or $\Delta t=240-400$~months, whence
justifying the normalization ''16'' as the safest lowest boundary range for quasi
Brownian fluctuations. Starting that far with a Gaussian as initial condition,
the results of the integration (for $\tau$ = 0,1,2,3,4, or $\Delta t$= 16, 8, 4,
2, 1) are rather trivial. The variance and the mean are found to vary very
weakly. The experimental data pdf's are shown in Fig. 6, - together even with the
pdf for $\Delta t=400$~months, - the latter being indistinguishable from the
$\Delta t=16$~months case. It is readily remarkable that the pdf's are close to
the Gaussian form in the interesting NAO ($\Delta x$) index range, - with some
marginal deviation for the extreme (and rare) events.

\subsection{Discussion}

It is therefore confirmed that the NAO is a complex phenomenon which is almost
Markovian. This stresses the need to insert appropriate feedback mechanisms into
any model evolution equation(s), with an appropriate (red) noise term. This has
been recently
discussed\cite{PenlandandMatrosova1998,PenlandandSardeshmukh1995,Johnson et al.
2000,Fernandez et al. 2003}. Long-range fractionally integrated noise seems
indeed to provide a better fit of the NAO SLP  wintertime index over the period
1864 -- 1998 than does either stationary red noise or a non-stationary random
walk\cite{Greatbatch 2000}.

The persistence of the NAO index fluctuations, i.e.  SLP fluctuations, is in
agreement with the persistence of the sea surface temperature fluctuations at
different sites in the North Atlantic as found by Monetti et al.\cite{Monetti et
al. 2003}. Some reasons for the above can be found in studies based on
circulation-like models\cite{Joseyetal2001}.

\section{Conclusions}

In summary, the aims of this paper are twofold : (i) to search whether scaling
ranges exist in the North Atlantic Oscillation pressure index ; (ii) to derive
its FPE and check the validity of the Markovianity assumption. This allows one to
examine different time scales on the same footing, - a fundamental need in
geophysics\cite{Fraedrich and Schonwiese 2002}.

It is found that the lack of departure from a Gaussian process definitely is a
new quality of the NAO index data set, - not perceptible with a rescaled range
analysis, or a DFA, as done above, nor with spectral studies.

However, it seems  that the actual NAO index is not very useful\footnote{This is
in agreement with conclusions from a recent paper casting doubt on the NAO-global
warming relationship\cite{Jonnson and Miles 2002}, and indicating a $too$
$strong$ influence of the Azores data with respect to the Iceland one. See also
Czaja and Frankignoul\cite{CzajaFrankignoul} }. Thus one might have to request
other measurements for better predictability of climate and weather in Europe and
the Northern Hemisphere, e.g. at other locations. We might also suggest studies
on ''not-monthly-averaged'' indices, -  a monthly average being strangely
unphysical in our opinion.

Yet it is  emphasized  that the FPE provides the complete knowledge as to how the
statistics of correlations in the index distribution change on different delay
times. Since this includes an analysis in time $t$ for a $scalar$ $\Delta x$, it
seems that the findings could be implemented in atmospheric weather low
dimensional $vector-models$\cite{Grabowski and Smolarkiewicz 1999,Ragwitz and
Kantz 2000}. Further work in line with the above should be to relate the FPE to
an analytical solution,e.g. with a model of the turbulence-like dynamics as was
done  for financial indices\cite{Ausloos and Ivanova 2003} or ionic transport
through membranes\cite{Ausloos et al. 2004} through a Beck-Tsallis\cite{Tsallis
1988,Beck 2001} nonextensive thermodynamics approach.

\vskip 1.6cm {\bf Acknowledgments} \vskip 0.6cm

We thank D. Stauffer, C. Nicolis, K. Ivanova, and A. Bunde,  for stimulating
discussions and comments. Email correspondence with J. Peinke, Ch. Renner and R.
Friedrich is specifically appreciated.

\newpage

{\bf Appendix }

The coefficients called $drift$ and $diffusion$ used in the main text pertain to
the evolution of the pdf; they are usually describing the motion of particles in
fluid mechanics. For example the diffusion coefficient occurs in the Einstein or
Langevin equation of Brownian motion, as

\begin{equation} D = \frac {k_B T}{6\pi \eta a} \label{b1} \end{equation} where
$T$ is the bath temperature, $\eta $ the dynamic viscosity of the fluid and $a$
the diameter of the particle, such that the evolution of the particle is
described as a function of time $t$ by $<x^2>$ = $2Dt$, solution of the Langevin
equation

\begin{equation} M \frac {d^2x}{dt^2} = -{6\pi \eta a} \frac {dx}{dt} + R(t),
\end{equation} where $M$ is the mass of the particle and $R(t)$ a random force
with zero mean.

The Langevin equation is equivalent to the standard diffusion equation for a
probability density

\begin{equation} \frac{d p(x,t)}{dt }= D \frac{\partial ^{2} p(x,t)}{\partial
x^{2}} \label{b2} \end{equation} for which the solution is a Gaussian

\begin{equation} p(x,t) = \frac{1}{\sqrt(4 \pi D t) } e^{-x^2/(4Dt)} . \label{b3}
\end{equation}

The FPE written in the main text contains an extra term to Eq.(14); let it be
rewritten here as

\begin{equation} \frac{d}{dt}p(x,t)=[-\frac{\partial }{\partial x}
D^{(1)}(x,t)+\frac{\partial }{\partial ^{2}x^{2}} D^{(2)}(x,t)]p(x,t), \label{b4}
\end{equation} from which the mean $<x>$ and the variance $<\sigma^2 > $  can be
defined as usual. Keeping only the linear term in $D^{(1)}$, with a coefficient
$D_1^{(1)}$ and the independent $D_{0}^{(2)}$ and quadratic $D_{2}^{(2)}$terms in
$D^{(2)}$, from Eq.(9), one easily obtains the evolution of the ''particle'' as

\begin{equation} \left\langle x(t)\right\rangle = \left\langle
x_0(t)\right\rangle e^{(2D_{1}^{(1)}+D_{2}^{(2)})t} \label{b5} \end{equation} and
a similar equation for the variance, but also containing $D_{0}^{(2)}$, from
which one observes that $D^{(1)}$ and $D^{(2)}$ are $true$ drift and diffusion
coefficients. Note the time scale given by the inverse of $D^{(1)}$ and
$D^{(2)}$, i.e., about 1 month.

Knowing that $ \Delta x$ is the NAO index, a difference in pressure, $(\Delta P)$
we can roughly rewrite the ''official'' diffusion coefficient, Eq.(12), as

\begin{equation} <(\Delta P)^2> = \frac { 2k_B T}{6\pi \eta a} t,\label{b6}
\end{equation} and ''interpret it'', suggesting that in further and more precise
work, one could develop a model relating the changes in pressure (between Iceland
and the Azores) with a temperature field (in principle a temperature gradient,
rather than the mean temperature of the bath).

No need to say that the solution of a Brownian particle in a (rotating) bath
under a temperature gradient and with a noise force term is indeed what a good
weather model is (or should be).

 \newpage

{\bf Figure Captions} \vskip0.5cm

\textbf{Figure 1} Time evolution of the monthly averaged NAO index fluctuations
from January 1825 to November 2002 (2135 points) available on the web sites
$http://www.cru.uea.ac.uk/$$ftpdata/nao.dat$ and updated at
$http://www.cru.uea.ac.uk/$$\tilde{}$$timo/projpages/nao_{-}update.htm$)
\vskip0.5cm

\textbf{Figure 2} Integration of the NAO index fluctuations signal shown on
Figure 1

\vskip0.5cm

\textbf{Figure 3} Rescaled range analysis of the NAO index signal. Inserts:
Rescaled range analysis of the NAO index signal shortened by 5\% at the beginning
of the data series (left upper panel) or at the end (right lower panel)

\vskip0.5cm

\textbf{Figure 4 } Detrended fluctuation function of the (integrated) NAO index
signal. Inserts: Detrended fluctuation function of the (integrated) NAO index
signal after shortening by 5\% at the beginning of the data series (left upper
panel) or at the end (right lower panel) \vskip0.5cm

\textbf{Figure 5 } Kramers-Moyal coefficients (a) $M^{(1)}$ and (b) $M^{(2)}$
estimated from the empirical conditional density probability of the NAO
distribution values. The solid curves represent a linear and a quadratic fit,
respectively, excluding large events

\vskip0.5cm

\textbf{Figure 6 } Raw data (symbols), and theoretical (Gaussian) pdf (solid
line) comparing the NAO fluctuation distribution functions for various time lags,
$\Delta t$ = 16, 8, 4, 2, 1 months (or for $\tau$ = 0,1,2,3,4). The case $\Delta
t$ = 400 months is also ''shown'' for comparison

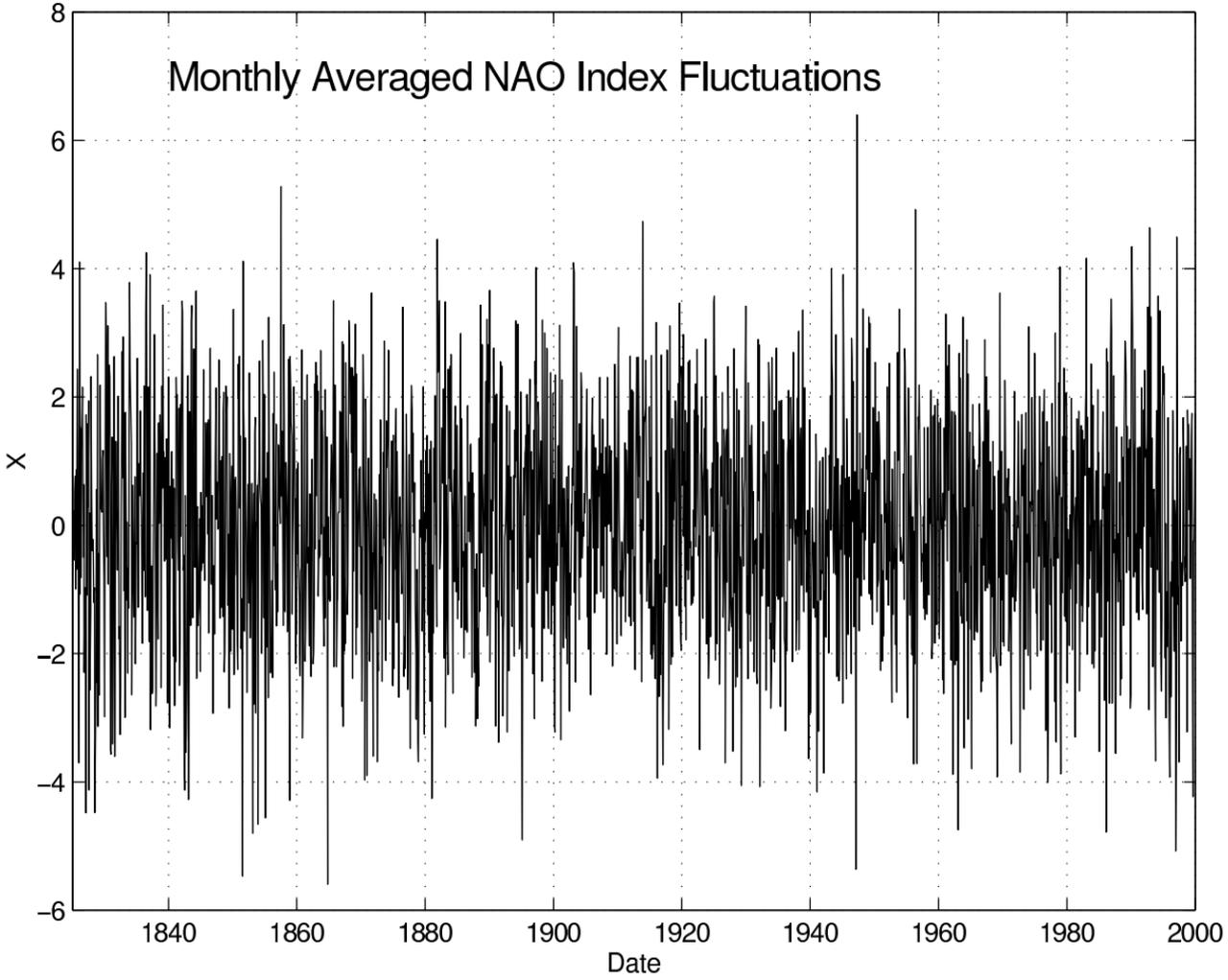
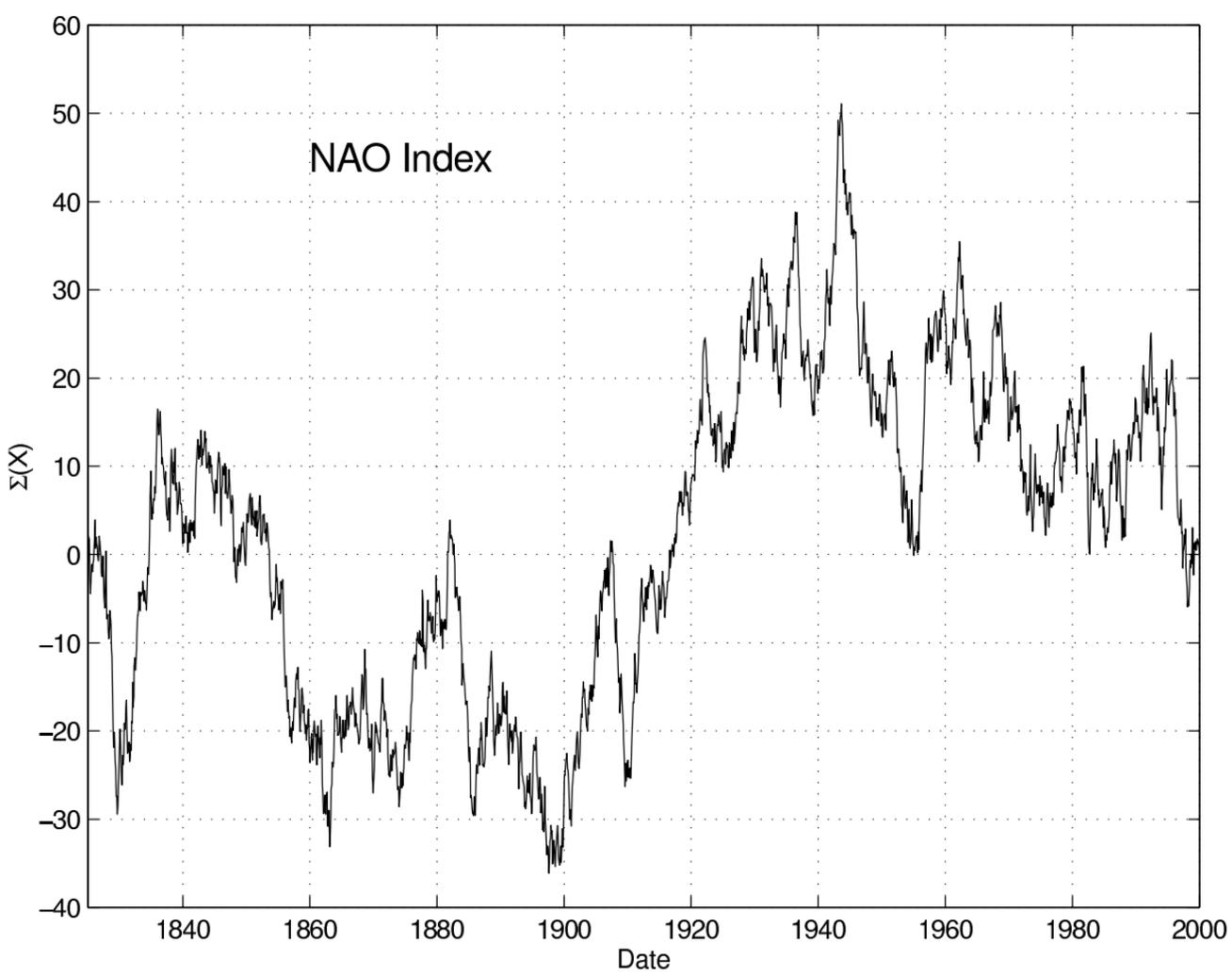
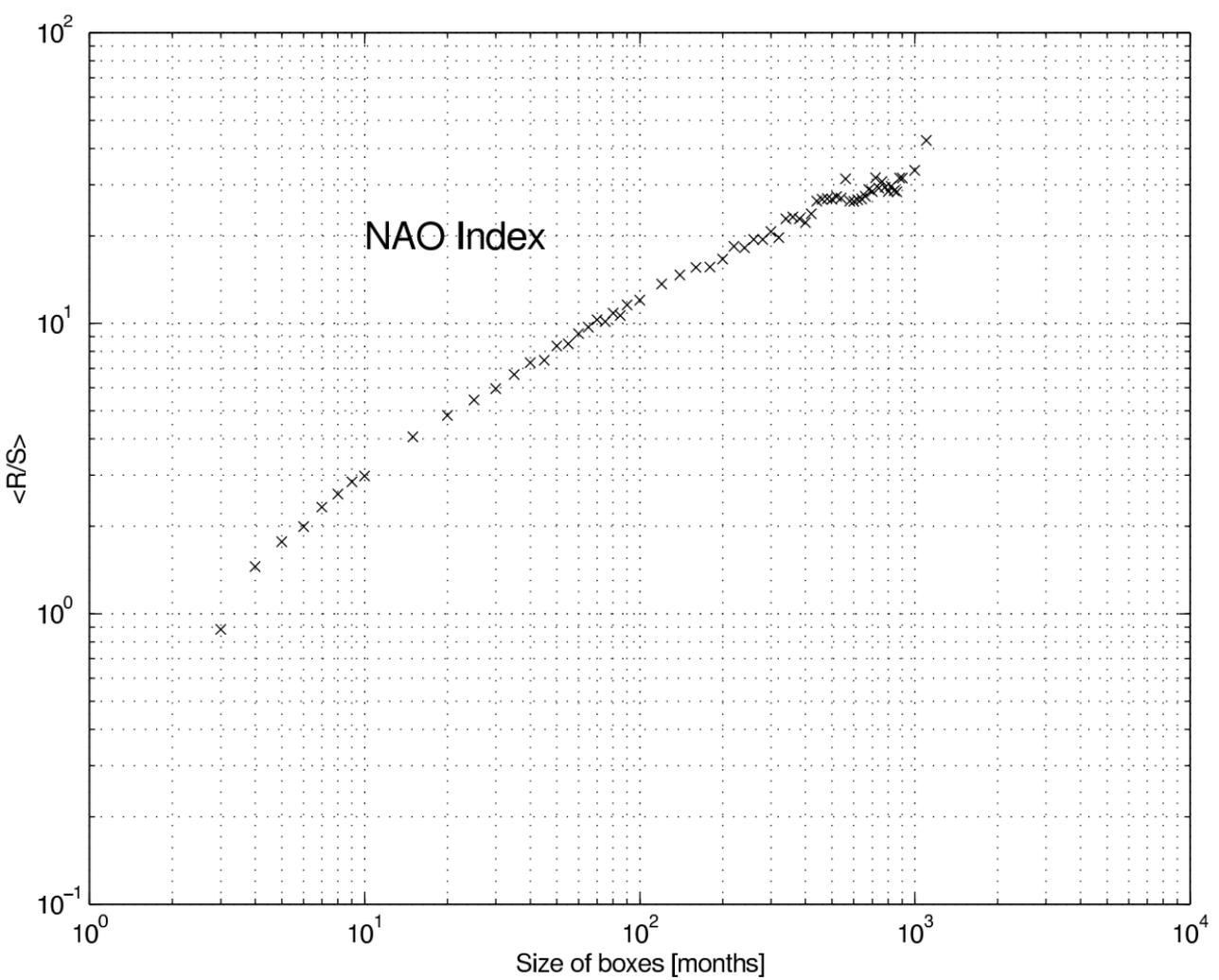
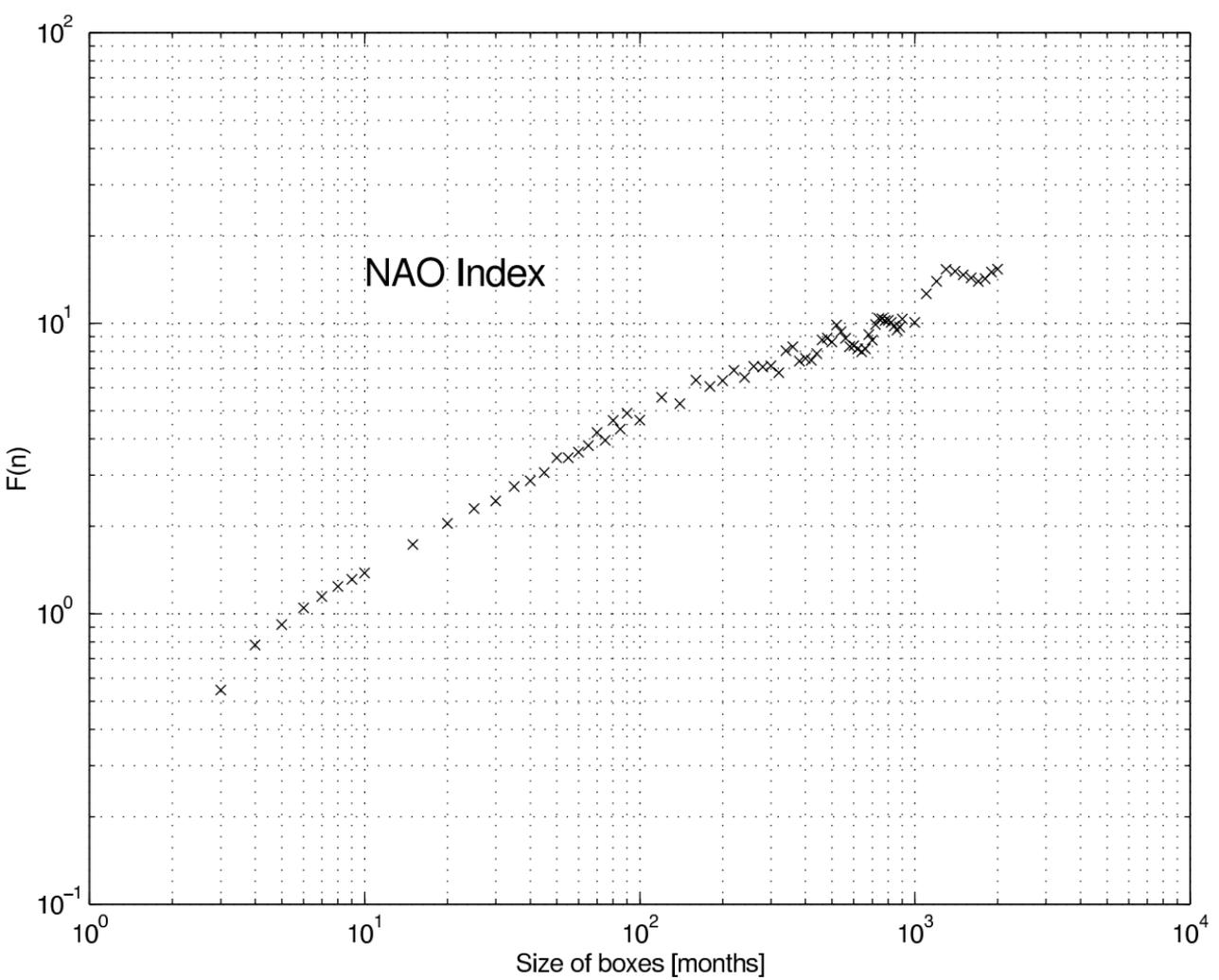
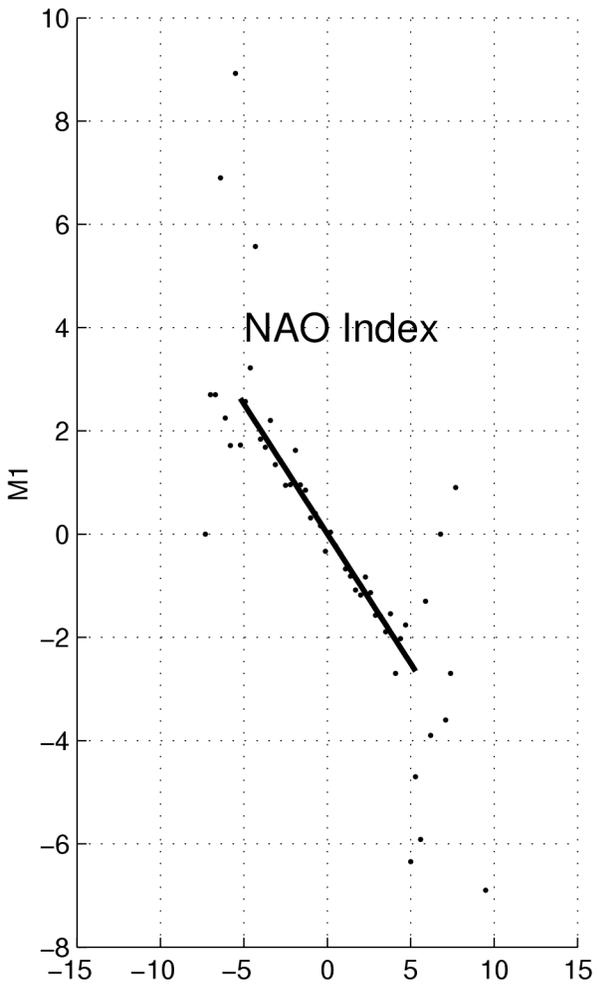
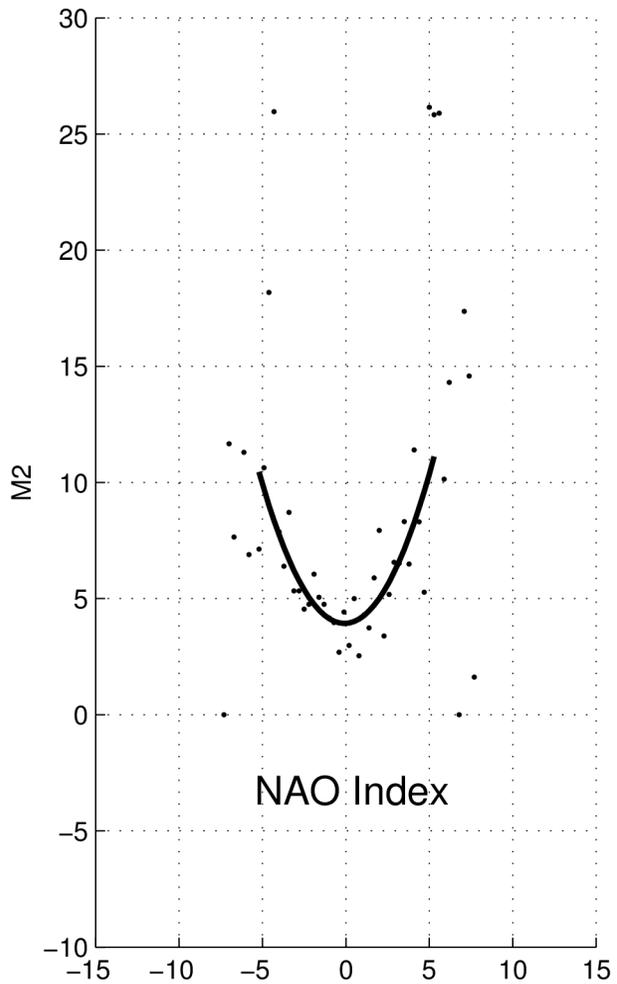
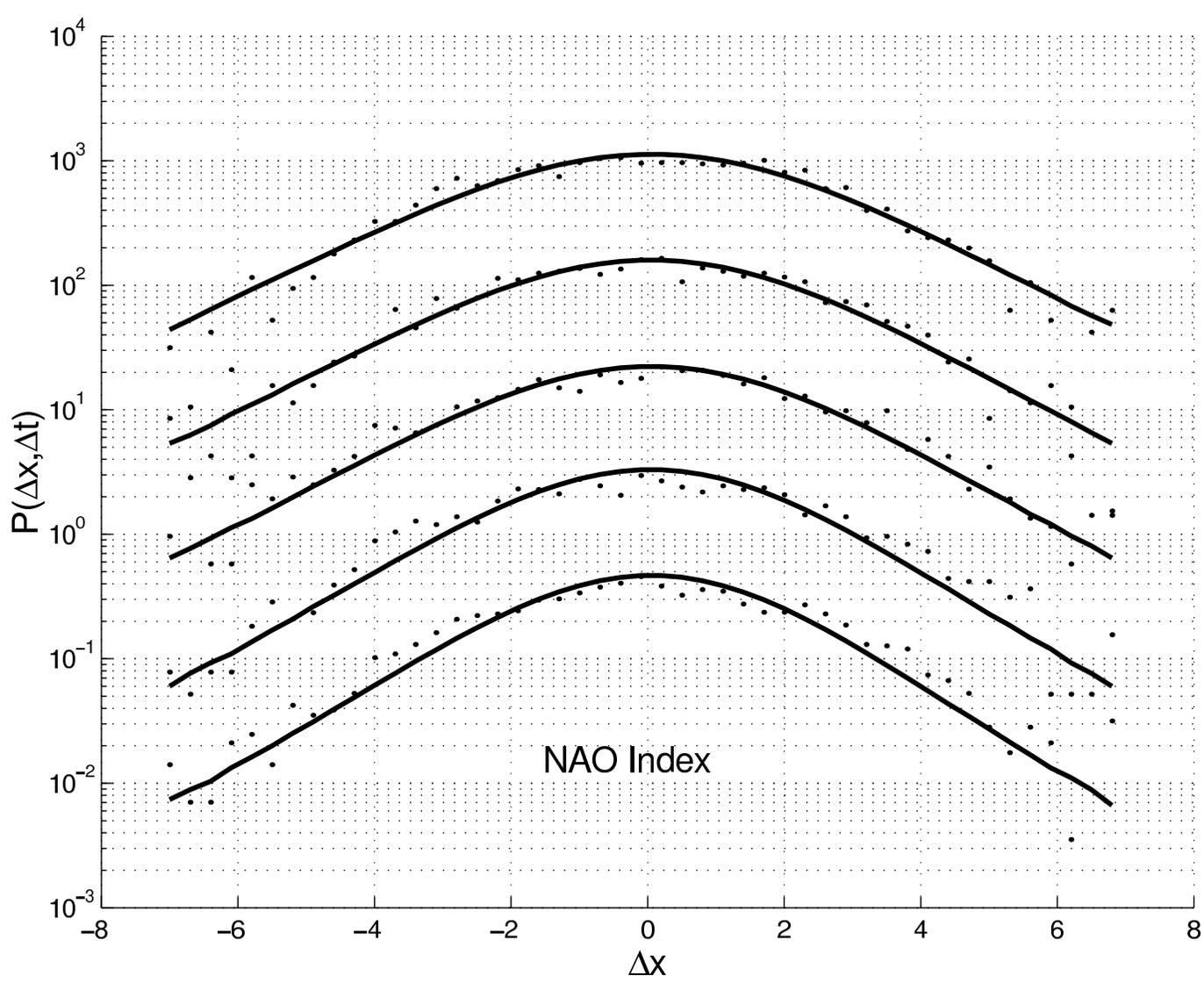
\end{document}